\documentclass[a4paper,11pt]{article}
\pdfoutput=1 % if your are submitting a pdflatex (i.e. if you have
\usepackage{jheppub} % for details on the use of the package, please
\usepackage{tikz}
\usepackage[T1]{fontenc} % if needed
\usetikzlibrary{shapes.misc}
\usetikzlibrary{decorations.pathreplacing,decorations.markings}

\tikzset{cross/.style={cross out, draw=black, minimum size=2*(#1-\pgflinewidth), inner sep=0pt, outer sep=0pt},
cross/.default={1pt}}
\tikzset{
  on each segment/.style={
    decorate,
    decoration={
      show path construction,
      moveto code={},
      lineto code={
        \path [#1]
        (\tikzinputsegmentfirst) -- (\tikzinputsegmentlast);
      },
      curveto code={
        \path [#1] (\tikzinputsegmentfirst)
        .. controls
        (\tikzinputsegmentsupporta) and (\tikzinputsegmentsupportb)
        ..
        (\tikzinputsegmentlast);
      },
      closepath code={
        \path [#1]
        (\tikzinputsegmentfirst) -- (\tikzinputsegmentlast);
      },
    },
  },
  mid arrow/.style={postaction={decorate,decoration={
        markings,
        mark=at position .5 with {\arrow[#1]{stealth}}
      }}},
}

\title{\boldmath The High Energy Behavior of Mellin Amplitudes}

\author[a]{Matthew Dodelson}
\author[a,b]{and Hirosi Ooguri}

\affiliation[a]{Kavli Institute for the Physics and Mathematics of the Universe (WPI),\\The University of Tokyo, Kashiwa, 277-8583, Japan}
\affiliation[b]{Walter Burke Institute for Theoretical Physics,\\
California Institute of Technology, Pasadena, CA 91125, USA}

\emailAdd{matthew.dodelson@ipmu.jp}
\emailAdd{hirosi.ooguri@ipmu.jp}

\abstract{In any consistent massive quantum field theory there are well known bounds on scattering amplitudes at high energies. In conformal field theory there is no scattering amplitude, but the Mellin amplitude is a well defined object analogous to the scattering amplitude. We prove bounds at high energies on Mellin amplitudes in conformal field theories, valid under certain technical assumptions. Such bounds are derived by demanding the absence of spurious singularities in position space correlators. We also conjecture a stronger bound, based on evidence from several explicit examples.}

\begin{document} 
\begin{flushright}
CALT-TH 2019-049\\
IPMU19-0160
\end{flushright}
\maketitle

\flushbottom

\section{Introduction}
In a field theory where the scattering amplitude is well-defined, it is interesting to study scattering at high energies. In this regime the scattering amplitude often simplifies, even in strongly coupled theories. In fact, there are universal bounds on scattering amplitudes and cross-sections at high energies. One example is the Froissart bound on the forward amplitude in a four-dimensional field theory \cite{froissart}, 
\begin{align}
A(\theta=0,E)<E^2\log^2 E\text{ as }E\to \infty.\label{FroissartBound}
\end{align}
This follows from the general principles of unitarity, assuming that there are no massless particles in the theory.\\
\indent  In a conformal field theory it is not immediately clear how to derive analogous bounds, since there is no well-defined scattering amplitude. One can use the conformal bootstrap to place bounds on the operator spectrum or the OPE coefficients, but these objects are not directly related to high energy dynamics. In this work we will instead derive high energy bounds on the Mellin amplitude, which plays the role of the scattering amplitude for conformal theories \cite{mack,penedones,fitzpatrick1,fitzpatrick2,TASI}. The bounds will hold for any conformal field theory, and do not rely on an expansion in $1/N$ or the coupling constants of the theory. \\
\indent The basic assumption that we will make is that the only singularities of correlation functions in Lorentzian signature arise from Landau diagrams, which are Feynman diagrams with lightlike lines conserving momentum at each vertex. This assumption has been proven in (1+1)-dimensional CFT \cite{looking}, but has not been shown in higher dimensions. For theories which satisfy this assumption, we can derive the
following bound on the $(d+2)$-point Mellin amplitude for conformal field theory in $d$ dimensions,
\begin{align}
\frac{M_{d+2}(\lambda^2 \gamma_{ij})}{\lambda^{d+1 - (d+2) \Delta}}\to 0\text{ as $\lambda^2\to -i\infty$ with $\gamma_{ij}$ fixed, physical, and generic,}
\end{align}
where  the Mellin amplitude $M_{d+2}(\gamma_{ij})$ is defined in (\ref{Mellin}) and
$\Delta$ is the conformal weight of the $d+2$ operators, which we assume to be the same.
The strategy is to show that if the Mellin amplitude did not fall sufficiently fast at high energies, then there would be a singularity in the $(d+2)$-point correlation function for which it is not possible to draw a Landau diagram. 
\\
\indent The structure of the paper is as follows. In Section 2 we review the Mellin formalism. In Section 3 we review the structure of Landau diagrams in 1+1 and generalize it to higher dimensions. 
In Section 4 we place a bound on the four point amplitude in 1+1 dimensions, which is then generalized to higher dimensions in Section 5. In Section 6 we specialize to the case of a holographic theory, and show that the bounds are satisfied once 
stringy effects in the bulk are taken into account. In Section 7 we will conjecture a stronger form of the bound, based on several examples. We conclude in Section 8 with some open questions. 
\section{Review of Mellin Variables}
We are used to analyzing CFT correlators in momentum space or position space. Mellin space \cite{mack,penedones,fitzpatrick1,fitzpatrick2,TASI} is a third way of describing CFT correlation functions. It is convenient because the Mellin variables are more closely related to Mandelstam variables in quantum field theory, and therefore the scattering matrix. In this section we will review some of the basic definitions. \\
\indent The definition of the $n$-point Mellin amplitude $M_n$ is
\begin{align}
\left\langle \prod_{i=1}^{n} \mathcal{O}_i(x_i)\right\rangle=\int_{c-i\infty}^{c+i\infty} \prod d\gamma_{ij}\,M_n(\gamma_{ij})\prod_{i<j=1}^{n}\frac{\Gamma(\gamma_{ij})}{(x_{ij}^2)^{\gamma_{ij}}}.
\label{Mellin}
\end{align}
The contour runs parallel to the imaginary axis, to the right of the poles in the $\Gamma$ functions, and it only runs over the independent Mellin variables. Here the matrix of Mellin variables $\gamma_{ij}$ can be taken to be symmetric, satisfying the constraint 
\begin{align}
\sum_{j\not=i}^{n}\gamma_{ij}=\Delta_i.
\end{align} To solve this constraint, we can introduce $n$ Lorentzian momenta $k_i$ with $\sum_{i=1}^{n}k_i=0$ and $k_i^2=-\Delta_i$. These conditions are analogous to momentum conservation and the on-shell condition. Then we can take $\gamma_{ij}=k_i\cdot k_j$.
So we see that the Mellin variables are parameterized by on-shell momenta satisfying momentum conservation, just like the scattering matrix. For $n\le d+2$, the number of independent Mellin variables is $n(n-3)/2$, which matches the number of independent cross ratios.  For $n>d+2$ the counting of variables is more complicated \cite{TASI}, but we will not deal with such cases here. \\
\indent For example, let's consider the case $n=4$, where for the scattering matrix we have kinematic invariants, {\it i.e.}, the Mandelstam variables. These correspond to independent Mellin variables $\gamma_{12}$ and $\gamma_{14}$, which are like $-s/2$ and $-u/2$ respectively. Plugging in and setting all the $\Delta$'s equal, we get the four-point function
\begin{align}
&\frac{1}{((x_{13})^2(x_{24})^2)^{\Delta}}\int_{c-i\infty}^{c+i\infty}\frac{d\gamma_{12}}{2\pi i}\int_{c-i\infty}^{c-i\infty}\frac{d\gamma_{14}}{2\pi i}\,M_4(\gamma_{ij})\Gamma(\gamma_{12})^2\Gamma(\gamma_{14})^2\notag\\
&\hspace{80 mm}\Gamma(\Delta-\gamma_{12}-\gamma_{14})^2|z|^{-2\gamma_{12}}|1-z|^{-2\gamma_{14}},
\end{align}
where $z$ is the usual cross-ratio. There are two kinds of infinite sequences of integer-spaced poles in the integrand for $\gamma_{12}$, one ascending and one descending. The constant $c$ should be chosen to be in the middle of these two sequences. \\
\indent The definitions so far are independent of holography, but they are particularly simple for the case of a theory with a local bulk. Contact interactions in the bulk give polynomial Mellin amplitudes, just like contact interactions in flat space give polynomial scattering amplitudes. An exchange of a spin-$l$ field with dimension $\Delta$ gives
\begin{align}
\sum_{m=0}^{\infty}\frac{Q_{l,m}(t)}{s-\Delta+l-m}+R(s,t),
\end{align}
where $R$ is analytic and $Q$ is a polynomial. There is an infinite sequence of poles at integer spaced values, corresponding to the exchange of descendants in the $s$ channel.
\section{Singularities in Lorentzian CFT}

\subsection{The four point function in 1+1 dimensions} 
\indent In this subsection we will review the kinematics in which the bulk point singularity emerges \cite{looking,gary}. We work in complex coordinates on the Euclidean plane, starting in the following Euclidean configuration:
\begin{align}
(z_1,z_2,z_3,z_4)=(-\rho,\rho,1,-1),
\end{align}
where $\rho=e^{i\phi-\epsilon}$, $0<\phi<\pi$, and $\epsilon>0$. Now we increase the Lorentzian time $\tau_L$, so that $(\rho,\overline{\rho})=(e^{i(\tau_L+\phi)-\epsilon},e^{i(\tau_L-\phi)-\epsilon})$. Then the cross-ratios are
\begin{align}
z&=\frac{z_{12}z_{34}}{z_{13}z_{24}}=\frac{4\rho}{(1+\rho)^2}=\sec^2\left(\frac{\tau_L+\phi+i\epsilon}{2}\right)\notag\\\overline{z}&=\frac{\overline{z}_{12}\overline{z}_{34}}{\overline{z}_{13}\overline{z}_{24}}=\frac{4\overline{\rho}}{(1+\overline{\rho})^2}=\sec^2\left(\frac{\tau_L-\phi+i\epsilon}{2}\right).
\end{align}
When $\epsilon\to 0$ we have $z,\overline{z}>1$. Note that $z=\overline{z}$ both at $\tau_L=0$ and $\tau_L=\pi$:
\begin{align}
z(\tau_L=0)&=\overline{z}(\tau_L=0)=\sec^2\left(\frac{\phi}{2}\right)\notag\\
z(\tau_L=\pi)&=\overline{z}(\tau_L=\pi)=\csc^2\left(\frac{\phi}{2}\right).
\end{align}
\indent We are interested in what happens as we increase $\tau_L$ from zero to $\pi$, the latter being the location of the bulk Landau singularity. 
Along the way we run into several interesting points. When $\tau_L=\phi$ we have $\overline{z}-1=-\epsilon^2/4$, and when $\tau_L=\pi-\phi$ we have $z=-4/\epsilon^2$. It follows that $z$ traces a contour in the complex plane that goes counterclockwise around the points $z=0$ and $z=1$, and $\overline{z}$ traces a contour in the complex plane that goes clockwise around just the point $\overline{z}=1$. 

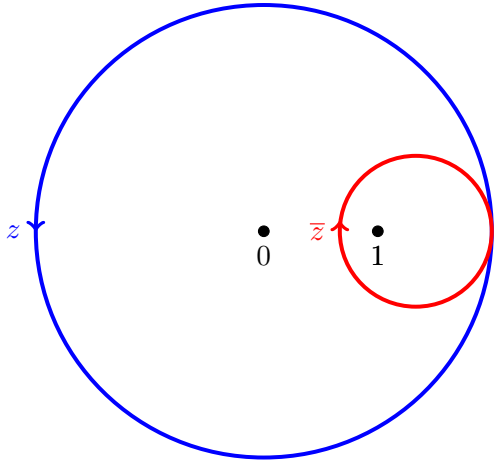
\begin{figure}
\begin{center}\begin{tikzpicture}
\draw[
    blue, line width=1.5,   decoration={markings, mark=at position 0.5 with {\arrow{>}}},
        postaction={decorate}
        ]
        (0,0) circle (3);
    \draw[ red, line width=1.5,
        decoration={markings, mark=at position .5 with {\arrow{<}}},
        postaction={decorate}
        ]
        (2,0) circle (1);
\filldraw (0,0) circle (2pt);
\filldraw (1.5,0) circle (2pt);

    \node[below, outer sep=2pt] at (1.5,0) {1};
        \node[blue,left, outer sep=2pt] at (-3,0) {$z$};
                \node[red,left, outer sep=2pt] at (1,0) {$\overline{z}$};
    \node[below, outer sep=2pt] at (1.5,0) {1};
        \node[below, outer sep=2pt] at (0,0) {0};
\end{tikzpicture}
\end{center}
\caption{The contours traced out by $z$ and $\overline{z}$ along the analytic continuation from $\tau_L=0$ to $\tau_L=\pi$. For simplicity we have taken $\phi=\pi/2$, so that $z(\tau=0)=z(\tau=\pi)$.}
\end{figure}
\indent It is known that the bulk point singularity at $z=\overline{z}$ must be resolved in a consistent CFT in 1+1 dimensions \cite{looking}. This will be our starting point for deriving the main result in this paper. 
\subsection{The $(d+2)$ point function in $d$-dimensions}
\indent In this subsection we will generalize the previous analysis to arbitrary dimensions. 
Let us start with the case of $d=3$. 
Following \cite{looking}, we use the embedding space defined by 
\begin{align}
-(X^1)^2-(X^2)^2+(X^3)^2+(X^4)^2+(X^5)^2=0.
\end{align}
Let us choose the $5$ points to be located at
\begin{align}
X_1&=(1,0,1,0,0)\notag\\
X_2&=(1,0,-1,0,0)\notag\\
X_3&=(\cos\tau_L,\sin\tau_L,0,-1,0)\notag\\
X_4&=(\cos\tau_L,\sin\tau_L,-\cos\theta_4,-\sin\theta_4\cos\phi_4,-\sin\theta_4\sin\phi_4)\notag\\
X_5&=(\cos\tau_L,\sin\tau_L,-\cos\theta_5,-\sin\theta_5\cos\phi_5,-\sin\theta_5\sin\phi_5).
\end{align}
We can assemble them into a $5 \times 5$ matrix $X$. 
In this language the bulk point singularity exists when $\det X=0$. \\
\indent The rank of the matrix $X$ is an interesting invariant. We have $\text{Rank }X=4$ at $\tau_L=\pi$ for generic angles, and $\text{Rank }X=3$ at $\tau_L=\pi$ and $\sin\phi_4=\sin\phi_5=0$ (with generic $\theta_4$ and $\theta_5$). In the rank 3 case all five points lie on a great circle. The determinant is 
\begin{align}
\det X=\sin\tau_L(\sin\theta_4\sin\phi_4-\sin\theta_5\sin\phi_5-\sin\theta_4\sin\theta_5\sin(\phi_4-\phi_5)).
\end{align}
The first factor vanishes in the rank 4 case, and both factors vanish in the rank 3 case.\\
\indent The claim is that if there exists a boundary Landau diagram at $\tau_L=\pi$, then $\text{Rank }X<4$. Consider a Landau diagram at $\tau_L=\pi$. $x_1$ and $x_2$ shoot out light rays, which collide at the equator at time $\pi/2$. In order to match up with the final points at time $\pi$, the final points must be a distance $\pi/2$ away from the interaction point. This implies that $\sin\phi_4$ and $\sin\phi_5$ are both zero. In other words, all five points are on a great circle. Thus we have $\text{Rank }X<4$.\\
\indent There is one more necessary condition for the existence of a Landau diagram. Momentum must be conserved at the interaction point, which means that the initial two points cannot be adjacent on the great circle. In other words at least one of the angles $\phi_4$ and $\phi_5$ must equal $\pi$. \\
\indent This result generalizes to higher dimensions: if there is a boundary Landau diagram in the $(d+2)$-point function at $\tau_L=\pi$, then $\text{Rank }X<d+1$. Conversely if $\text{Rank }X<d+1$ then a boundary Landau diagram exists, provided momentum conservation can be satisfied at the vertex.
\begin{figure}
\begin{center}\begin{tikzpicture}
  \shade[ball color = gray!40, opacity = 0.4] (0,0) circle (2cm);
  \draw (0,0) circle (2cm);
  \draw (-2,0) arc (180:360:2 and 0.6);
  \draw[dashed] (2,0) arc (0:180:2 and 0.6);
\filldraw [red] (0,2) circle (2pt)
(0,-2) circle (2pt);
\filldraw [blue] (2,0) circle (2pt)
(-2*.86,2*1/2) circle (2pt)
(2*1/2,-2*.86) circle (2pt);
\draw (0,-.6) node[cross=3pt] {};
\end{tikzpicture}
\hspace{30 mm}
\begin{tikzpicture}
  \shade[ball color = gray!40, opacity = 0.4] (0,0) circle (2cm);
  \draw (0,0) circle (2cm);
  \draw (-2,0) arc (180:360:2 and 0.6);
  \draw[dashed] (2,0) arc (0:180:2 and 0.6);
\filldraw [red] (0,2) circle (2pt)
(0,-2) circle (2pt);
\filldraw [blue] (2,0) circle (2pt)
(2*.86,2*1/2) circle (2pt)
(2*1/2,-2*.86) circle (2pt);
\draw (0,-.6) node[cross=3pt] {};
\end{tikzpicture}
\end{center}
\caption{Two examples of configurations of five points on a great circle on $S^2$. The red points are at $\tau=0$, and the blue points are at $\tau=\pi$. The interaction vertex is marked with a cross. In the left hand configuration it is possible to conserve momenta at the vertex, so there is a boundary Landau singularity. In the right hand configuration there is no way to conserve horizontal momentum at the vertex, so there is no boundary Landau diagram.}
\end{figure}
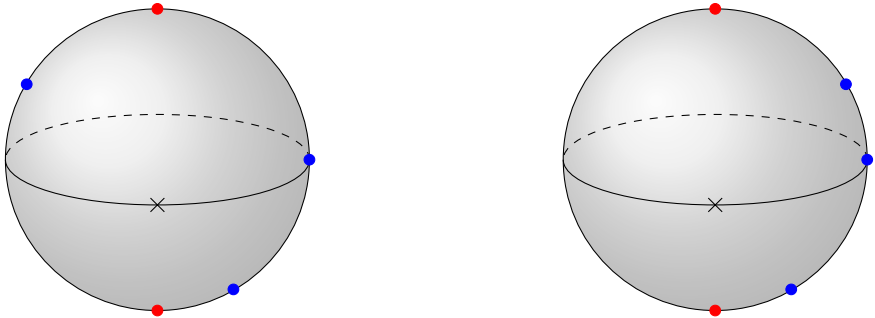\\
\indent Not much is known about which Lorentzian singularities arise in a generic CFT in more than 1+1 dimensions. In a perturbative theory the only allowed singularities come from Landau diagrams, but additional singularities could arise in strongly coupled theories. In particular it is not necessarily true that the bulk point singularity will be generically resolved, although we are not aware of any explicit examples in which this singularity remains. In this paper we will derive bounds on the Mellin amplitude in theories for which the bulk point singularity is resolved. 
\section{Bounding the Four Point Amplitude in 1+1 Dimensions}
\indent We consider the four point Mellin amplitude in 1+1 dimensions. The basic dynamical information that we will use to bound the Mellin amplitude is the fact that there is no singularity at $z=\overline{z}$. So we first need to understand what regime of Mellin space dominates near $z=\overline{z}$. The claim is that this is the hard scattering regime, or in other words the regime of large Mellin variables with all ratios of Mellin variables fixed \cite{penedones}. \\
\indent Let's prove this claim by examining the contribution of the hard scattering regime to the four-point function. Stirling's approximation for $|\gamma_{12}|,|\gamma_{14}|\gg \Delta$ and fixed $\gamma_{12}/\gamma_{14}$ gives 
\begin{align}
&e^{-i\Delta\tau}|z|^{2\Delta}\int_{c-i\infty}^{c+i\infty}d\gamma_{12}\int_{c-i\infty}^{c-i\infty}d\gamma_{14}\, M_4(\gamma_{ij})\frac{(\gamma_{12}+\gamma_{14})^{2\Delta-1}}{\gamma_{12}\gamma_{14}}\notag\\
&\hspace{30 mm}\gamma_{12}^{2\gamma_{12}}\gamma_{14}^{2\gamma_{14}}(-\gamma_{12}-\gamma_{14})^{-2\gamma_{12}-2\gamma_{14}}|z|^{-2\gamma_{12}}|1-z|^{-2\gamma_{14}}
\end{align}
Because of the branch cuts in the integrand, we have been careful about not writing $(xy)^a=x^ay^a$ if $x,y$, and $a$ are complex numbers. Now we want to switch to angular variables. To do this we write the auxiliary momenta in center of mass frame:
\begin{align}
k_1&=(E,E,0)\notag\\
k_2&=(E,-E,0)\notag\\
k_3&=(-E,-E\cos\theta,-E\sin\theta)\notag\\
k_4&=(-E,E\cos\theta,E\sin\theta).
\end{align}
Then 
\begin{align}
\gamma_{12}=-2E^2,\hspace{10 mm}\gamma_{14}=2E^2\cos^2(\theta/2).
\end{align}
The Jacobian is $E^3\sin\theta$, so the integral becomes 
\begin{align}
&e^{-i\Delta \tau_L}|z|^{2\Delta}\int_{E^2=c-i\infty}^{E^2=c+i\infty} dE\,\int \frac{d\theta}{\sin\theta}\, M_4(E,\theta)E^{4\Delta-3}(\sin^2(\theta/2))^{2\Delta}\\
&\hspace{-5 mm}\times \exp\left(4E^2\left(-\log (-E^2)+\log(E^2)+\log\sin^2(\theta/2)+\cos^2(\theta/2)\log\cot^2(\theta/2)+\log|z|-\cos^2(\theta/2)\log|1-z|\right)\right).\notag
\end{align}
\indent We now want to look for a saddle point in $\theta$.  This requires that we make the technical assumption that $M_4$ doesn't vary too strongly with $\theta$ as $|E|\to \infty$. In particular
\begin{align}
\frac{1}{E^2}\frac{\partial \log M_4(E,\theta)}{\partial \theta}\to 0\text{ as }|E|\to \infty \text{ with }\theta\text{ fixed}.\label{assumption}
\end{align}
Then the saddle point equation for $\theta$ is 
\begin{align}
\log(|1-z|\tan^2(\theta/2)).
\end{align}
The solution is
\begin{align}
\theta(z)=\pm 2\arctan\left(\frac{1}{\sqrt{|1-z|}}\right).
\end{align}
This has a simple interpretation in terms of the boundary kinematics as we approach $\tau_L=\pi$. Plugging in the values of $z,\overline{z}$ after the analytic continuation to $\tau_L=\pi$, we find $\theta=\pm \phi$. In other words the scattering angle determined by the Mellin variables is equal to the angle which defines the boundary kinematics. \\
\indent We can now perform the $\theta$ integral using the saddle point approximation. This gives
\begin{align}
&\frac{(1+|1-z|)^{1-2\Delta}e^{-i\Delta \tau_L}|z|^{2\Delta}}{\sqrt{|1-z|}}\int_{E^2=c-i\infty}^{E^2=c+i\infty} \frac{dE}{|E|}\, M_4(E,\theta(z))\notag\\
&\hspace{10 mm}\times E^{4\Delta-3}\exp\left(4E^2\left(-\log(-E^2)+\log(E^2)+\log|z|-\log(1+|1-z|)\right)\right).
\end{align}
As a first step, let's consider the integrand at $\tau=0$. For the upper half of the contour, $E^2=i|E|^2$, so $\log(E^2)-\log(-E^2)=i\pi$. It follows that the integrand is exponentially damped as $e^{-4\pi|E|^2}$. Similarly, for the lower half of the contour, $\log(E^2)-\log(-E^2)=-i\pi$, so the integrand is again exponentially damped as $e^{-4\pi |E|^2}$. Therefore the integral converges, assuming a very weak condition on the Mellin amplitude:
\begin{align}
M_4(E,\theta(z))\exp(-4\pi |E|^2)\to 0\text{ as }|E|\to \infty.
\end{align}  
\indent What happens after the analytic continuation to $\tau=\pi$? The analytic continuation takes $\log|z|\to \log|z|+i\pi$. So the upper half of the contour is now damped as $e^{-4\pi |E|^2}$, but the lower half of the contour has no damping factor. The contribution of the lower half of the contour to the integral is then \begin{align}
&\frac{(1+|1-z|)^{1-2\Delta}e^{-i\Delta \tau_L}|z|^{2\Delta}}{\sqrt{|1-z|}}\int_{E^2=-i|E^2|} \frac{dE}{|E|}\, E^{4\Delta-3}M_4(E,\theta(z))\left(\frac{|z|}{1+|1-z|}\right)^{4E^2},\notag\\
&=\frac{z}{\sqrt{z-1}}\int_{E^2=-i|E^2|} \frac{dE}{|E|}\, M_4(E,\theta(z)) E^{4\Delta-3}\exp\left(\frac{E^2(z-\overline{z})^2}{2z^2(z-1)}\right).
\end{align}
where we took the $z\to \overline{z}$ limit. For a given value of $z-\overline{z}$, this integral is cut off at $E\sim1/(z-\overline{z})$. This means that as we take $z\to \overline{z}$, the integral is in danger of diverging. But we know that the bulk point singularity must be absent in the CFT, so the Mellin amplitude must fall fast enough to ensure convergence of the integral. This is the origin of the desired bound.\\
\indent To see how this works, let's first consider the example of a constant Mellin amplitude, which arises from a bulk contact interaction. Performing the integral, we find
\begin{align}
\frac{z^{4\Delta-2}(z-1)^{2(\Delta-1)}}{(z-\overline{z})^{4\Delta-3}}
\end{align}
This is singular at $z=\overline{z}$ if $\Delta>3/4$. It is much simpler in $(\tau_L,\phi)$ variables: 
\begin{align}
\frac{1}{\sin\phi}\frac{1}{(\tau_L-\pi)^{4\Delta-3}}. 
\end{align}
The singularity is most prominent near forward and backward scattering. As $\phi\to0$ it just becomes the light-cone singularity.\\
\indent More generally $M_4$ will not be constant. In order for the singularity to be resolved we need 
\begin{align}
\frac{M_4(E,\theta)}{E^{3-4\Delta}}\to 0\text{ as $E^2\to -i\infty$ with $\theta$ real, generic, and fixed.}
\end{align}
This is the main result of this paper. It relies only on the technical assumption (\ref{assumption}). In particular it is nonperturbative in $1/N$ and $1/\lambda$ in the case of a holographic theory. The bound can be easily generalized to nonidentical external dimensions, with the result that $4\Delta$ is replaced by $\sum_i \Delta_i$.
\section{A Bound in Higher Dimensions}
We now consider the $(d+2)$-point function in $d$ dimensions. The strategy is similar to the previous section. We have $d+1$ angular variables and one overall energy variable. We will look for saddles in the angular variables, and reduce the integral over Mellin space to an integral over the energy. Then we will find the condition on $M_{d+2}$ such that the bulk point singularity is resolved. The only difference is that it is not known for $d>2$ whether the bulk point singularity is resolved in a generic CFT. \\
\indent To be explicit, let's consider the five point function in $d=3$. We use $\gamma_{12},\gamma_{13},\gamma_{24},\gamma_{35}$, and $\gamma_{45}$ as the independent variables. After applying Stirling's approximation, we find
\begin{align}
&\int \frac{d\gamma_{12}\, d\gamma_{13}\, d\gamma_{24}\,d\gamma_{35}\, d\gamma_{45}}{\sqrt{\gamma_{12}\gamma_{13}\gamma_{24}\gamma_{35}\gamma_{45}}}\,M_5(\gamma_{ij})(\gamma_{35}-\gamma_{12}-\gamma_{24})^{(\Delta-1)/2}(\gamma_{24}-\gamma_{13}-\gamma_{35})^{(\Delta-1)/2}(\gamma_{45}-\gamma_{12}-\gamma_{13})^{(\Delta-1)/2}\notag\\
&\hspace{5 mm}(\gamma_{13}-\gamma_{24}-\gamma_{45})^{(\Delta-1)/2}(\gamma_{12}-\gamma_{35}-\gamma_{45})^{(\Delta-1)/2}\exp(F(\gamma_{ij},x_{ij})).
\end{align}
We have dropped an overall $x_{ij}$-dependent prefactor since it does not affect the derivation of the bound. The function $F$ is defined by  
\begin{align}
&F(\gamma_{ij},x_{ij})\\
&=\gamma_{12}\left(\log\gamma_{12}+\log (\gamma_{12}-\gamma_{35}-\gamma_{45})-\log(\gamma_{35}-\gamma_{12}-\gamma_{24})-\log(\gamma_{45}-\gamma_{12}-\gamma_{13})-\log\left(\frac{x_{12}^2x_{34}^2}{x_{14}^2x_{23}^2}\right)-2\pi i \right)\notag\\
&+\gamma_{13}\left(\log\gamma_{13}+\log(\gamma_{13}-\gamma_{24}-\gamma_{45})-\log(\gamma_{24}-\gamma_{13}-\gamma_{35})-\log(\gamma_{45}-\gamma_{13}-\gamma_{12})-\log\left(\frac{x_{13}^2x_{25}^2}{x_{15}^2x_{23}^2}\right)\right)\notag\\
&+\gamma_{24}\left(\log \gamma_{24}+\log(\gamma_{24}-\gamma_{13}-\gamma_{35})-\log(\gamma_{35}-\gamma_{24}-\gamma_{12})-\log(\gamma_{13}-\gamma_{24}-\gamma_{45})-\log\left(\frac{x_{15}^2x_{24}^2}{x_{14}^2x_{25}^2}\right)\right)\notag\\
&+\gamma_{35}\left(\log\gamma_{35}+\log(\gamma_{35}-\gamma_{12}-\gamma_{24})-\log(\gamma_{24}-\gamma_{35}-\gamma_{13})-\log(\gamma_{12}-\gamma_{35}-\gamma_{45})-\log\left(\frac{x_{14}^2x_{35}^2}{x_{15}^2x_{34}^2}\right)\right)\notag\\
&+\gamma_{45}\left(\log \gamma_{45}+\log(\gamma_{45}-\gamma_{12}-\gamma_{13})-\log(\gamma_{13}-\gamma_{45}-\gamma_{24})-\log(\gamma_{12}-\gamma_{45}-\gamma_{35})-\log\left(\frac{x_{23}^2x_{45}^2}{x_{25}^2x_{34}^2}\right)\right)\notag.
\end{align}
We hold $\gamma_{12}$ fixed, while varying the rest of the variables. The saddle point equations are 
\begin{align}
\frac{\gamma_{13}(\gamma_{13}-\gamma_{24}-\gamma_{45})}{(\gamma_{13}+\gamma_{35}-\gamma_{24})(\gamma_{12}+\gamma_{13}-\gamma_{45})}&=\frac{x_{13}^2x_{25}^2}{x_{15}^2x_{23}^2}\notag\\
\frac{\gamma_{24}(\gamma_{24}-\gamma_{13}-\gamma_{35})}{(\gamma_{12}+\gamma_{24}-\gamma_{35})(\gamma_{24}+\gamma_{45}-\gamma_{13})}&=\frac{x_{15}^2x_{24}^2}{x_{14}^2x_{25}^2}\notag\\
\frac{\gamma_{35}(\gamma_{35}-\gamma_{12}-\gamma_{24})}{(\gamma_{13}+\gamma_{35}-\gamma_{24})(\gamma_{35}+\gamma_{45}-\gamma_{12})}&=\frac{x_{14}^2x_{35}^2}{x_{15}^2x_{34}^2}\notag\\
\frac{\gamma_{45}(\gamma_{12}+\gamma_{13}-\gamma_{45})}{(\gamma_{12}-\gamma_{35}-\gamma_{45})(\gamma_{24}+\gamma_{45}-\gamma_{13})}&=\frac{x_{23}^2x_{45}^2}{x_{25}^2x_{34}^2}.
\end{align}
These are valid as long as $M_5$ doesn't vary too strongly with $\theta_i$ and $\phi_i$, 
\begin{align}
\frac{1}{E^2}\frac{\partial \log M_5(E,\theta_i,\phi_i)}{\partial \theta_i}\to 0,\frac{1}{E^2}\frac{\partial \log M_5(E,\theta_i,\phi_i)}{\partial \phi_i}\to 0\text{ as }|E|\to \infty\text{ with }\theta_i,\phi_i\text{ fixed}.
\end{align}
\indent To prove the desired bound it is sufficient to solve these equations at the bulk point singularity. The solution is given by $\gamma_{ij}=k_i\cdot k_j$, where 
\begin{align}
k_1&=\eta_1(E,E,0,0)\notag\\
k_2&=\eta_2(E,-E,0,0)\notag\\
k_3&=\eta_3(-E,0,-E,0)\notag\\
k_4&=\eta_4(-E,-E\cos\theta_4,-E\sin\theta_4\cos\phi_4,-E\sin\theta_4\sin\phi_4)\notag\\
k_5&=\eta_5(-E,-E\cos\theta_5,-E\sin\theta_5\cos\phi_5,-E\sin\theta_5\sin\phi_5),
\end{align}
where 
\begin{align}
\eta_1&=\sin\theta_4\sin\theta_5\left(-\sin\phi_4\cot(\theta_5/2)+\sin\phi_5\cot(\theta_4/2)+\sin(\phi_4-\phi_5)\right)\notag\\
\eta_2&=\sin\theta_4\sin\theta_5\left(-\sin\phi_4\tan(\theta_5/2)-\sin\phi_5\tan(\theta_4/2)+\sin(\phi_4-\phi_5)\right)\notag\\
\eta_3&=2\sin\theta_4\sin\theta_5\sin(\phi_4-\phi_5)\notag\\
\eta_4&=2\sin\theta_5\sin\phi_5\notag\\
\eta_5&=-2\sin\theta_4\sin\phi_4\label{etai}.
\end{align}
Here we have assumed that momentum conservation can be satisfied in the bulk $2\to 3$ scattering process, which implies that all the $\eta_i$'s are positive. \\
\indent Plugging in this saddle at $\tau_L=\pi$ gives 
\begin{align}
 \int_{E^2=-i|E|^2} dE\,  E^{5(\Delta-1)}M_5(E, \theta_i,\phi_i),
\end{align}
times an overall function of $\theta_i$ and $\phi_i$ that isn't relevant for our purposes. We conclude that the bulk point singularity is resolved if and only if 
\begin{align}
\frac{M_5(E,\theta_i,\phi_i)}{E^{4-5\Delta}}\to 0\text{ as $E^2\to -i\infty$ with $\theta_i,\phi_i$ physical, generic, and fixed.}
\end{align}
Generic means that we are away from regions where any number of the $\eta_i$'s goes to zero. Physical means that momentum conservation can be satisfied in the bulk scattering process for the given choice of angles. Fixed means that we do not vary $\theta_i$ and $\phi_i$ as we take $E^2\to -i\infty$.\\
\indent The generalization to arbitrary dimensions can be obtained by simply counting factors of $E$ in the integrand. Denoting the angles by $\theta_i$, we find
\begin{align}
\frac{M_{d+2}(E,\theta_i)}{E^{d+1-(d+2)\Delta}}\to 0\text{ as $E^2\to -i\infty$ with $\theta_i$ physical, generic, and fixed.}\label{generalbound}
\end{align}
\section{Resolution of Singularities in String Theory}
In the previous sections we derived a bound on the Mellin amplitude at high energies. But now there seems to be a paradox: holographic theories at infinite $N$ and $\lambda$ do not satisfy the bound. Indeed, the simple case of a contact interaction in the bulk gives a polynomial Mellin amplitude, which is far larger than required. The resolution of this tension in a weakly coupled bulk theory is that we need to resum all the string corrections. Stringy resolutions of the bulk point singularity
was discussed in \cite{looking}, and we will quantify it using the Mellin representation. \\
\indent To understand the effect of strings on the Mellin amplitude, we need to express the Mellin amplitude in terms of the flat space scattering amplitude $A_n(E,\theta_i)$ at large $N$ and large $\lambda$ \cite{penedones,okuda,fitzpatrick3,gary}. In this limit we have the relation
\begin{align} 
M_n(E,\theta_i)&=\int_{0}^{\infty}d\beta \,\beta^{(n\Delta-d)/2-1}e^{-\beta}A_n(\sqrt{2\beta} E,\theta_i)\notag\\
&=E^{d-n\Delta}\int_{0}^{\infty}d\beta \,\beta^{(n\Delta-d)/2-1}e^{-\beta/E^2}A_n(\sqrt{2\beta} ,\theta_i).
\end{align}
If the integral over each term in the Taylor expansion of the $e^{-\beta/E^2}$ factor converges, then at high energies the Mellin amplitude behaves as  $E^{d-n\Delta}$. This requires that $A_n$ decays faster than any power of $E$ at $E=\infty$. Assuming this fast decay rate at infinity, we find that $M_{d+2}$ behaves as $E^{d-(d+2)\Delta}$ as $|E|\to \infty$, which satisfies (\ref{generalbound}) at high energies. \\
     \indent It's instructive to compute $M_n$ in string theory. We can approximate $A_n$ by the high energy answer if $\beta  |E|^2\gg 1/\alpha'  $. Let's assume that this is true, and then check the assumption. Away from the positive $E^2$ axis, the scattering amplitude in tree-level string theory behaves as \cite{grossmende}
\begin{align}
A_n(E,\theta_i)\sim e^{-\alpha' E^2f_n(\theta_i)}\hspace{10 mm}\text{ for }|E|^2\gg 1/\alpha.
\end{align}
The Mellin amplitude then becomes
\begin{align}
M_n(E,\theta_i)&=\int_{0}^{\infty} d\beta \,\beta^{(n\Delta-d)/2-1}e^{-\beta(1+2\alpha' E^2f_n(\theta_i))}\notag\\
&\sim (1+2\alpha' E^2f_n(\theta_i))^{(d-n\Delta)/2}.
\end{align}
For large $\Delta$, the integral is dominated by $\beta\sim \Delta /(1+2\alpha' E^2f_n(\theta_i))$. So our assumption holds if  $|\Delta \alpha' E^2/(1+2\alpha' E^2f_n(\theta_i))|\gg 1$. This is satisfied when $\Delta\gg 1$ and $|E|^2\gg 1/(\Delta \alpha' )$. Note that since $A$ decays faster than any power of $E$ at infinity, the Mellin amplitude behaves as $E^{d-n\Delta}$ at high energies. \\
\indent This calculation is valid for generic angles. Now let us ask what happens for nongeneric angles, when we approach the kinematic limit corresponding to the boundary Landau singularity. For example, let's consider the five point function in $d=3$. The claim is that the rank 4 singularity is resolved, but the singularity with rank less than 4 is not. We first take $\tau\to \pi$, and then go near $\sin\phi_4,\sin\phi_5=0$. The five point function at $\tau=\pi$ is \cite{looking}
\begin{align}
\prod_i \eta_i^{\Delta-1}\int_{0}^{\infty} d\omega\, \omega^{5(\Delta-1)}A_5(\omega k_i n_i),
\end{align}
where $n_i$ is $X_i$ with the second entry removed and the $\eta_i$'s are defined in (\ref{etai}). Now take $\sin\phi_4,\sin\phi_5\to 0$, with $\sin\phi_4/\sin\phi_5$ held fixed. We rescale $\omega\to \omega/\sin\phi_4$, so that $\omega k_in_i$ doesn't vanish as $\sin\phi_4\to 0$ with $\sin\phi_4/\sin\phi_5$ fixed. Then we get 
\begin{align}
\frac{\prod_i \eta_i^{\Delta-1}}{(\sin\phi_4)^{5\Delta-4}}\int_{0}^{\infty} d\omega\, \omega^{5(\Delta-1)}A_5(\omega k_i n_i/\sin\phi_4),
\end{align}
This diverges as $1/\sin \phi_4$ as we go towards the rank 3 case. \\
\indent The upshot is that the $\det X$ singularity is resolved only when there is no boundary Landau singularity, in other words in the generic rank 4 case.
\section{A stronger bound?}
In the previous section we showed that for holographic CFT's, the Mellin amplitude $M_{d+2}$ behaves as $E^{d-(d+2)\Delta}$ at high energies, which is smaller than the required behavior $E^{d+1-(d+2)\Delta}$ by one power of $E$. We now consider another example with the same behavior, namely the four point Mellin amplitude of the operator $\sigma$ in the critical Ising model in 1+1 dimensions. This operator has conformal dimension $\Delta=1/8$, and the Mellin amplitude is \cite{alday}
\begin{align}
M_4=\frac{\Gamma\left(2\gamma_{12}-\frac{1}{4}\right)\Gamma\left(2\gamma_{14}-\frac{1}{4}\right)\Gamma\left(-2(\gamma_{12}+\gamma_{14})\right)}{\Gamma(\gamma_{12})^2\Gamma(\gamma_{14})^2\Gamma(1/8-\gamma_{12}-\gamma_{14})^2}.
\end{align}
At high energies, this behaves as $E^{3/2}$, which is smaller than $E^{3-4\Delta}=E^{5/2}$ by one power of $E$. \\
\indent Based on these two examples, it is natural to conjecture a possible stronger form of the bound, 
\begin{align}
M_{d+2}(E,\theta_i)\le f(\theta_i)E^{d-(d+2)\Delta},
\end{align}
where $f(\theta_i)$ is a function of the angles $\theta_i$. It would be interesting to check this stronger bound in other explicit examples.
\section{Open questions}
The main result of this paper was (\ref{generalbound}), a bound on the $(d+2)$-point Mellin amplitude in $d$ dimensions. This bound is independent of holography, but we demonstrated that the bound is satisfied if we consider a holographic theory with a weakly coupled bulk dual including strings. \\
\indent One question that we have not addressed here is whether there is a general bound on the $n$-point Mellin amplitude in $d$ dimensions. Also, it would be interesting to understand the general structure of Landau diagrams at $n$ points on $\mathbf{R}\times S^{d-1}$.\\
\indent Another question is whether loops in the bulk ever play a role for resolving the bulk point singularity. In a weakly coupled bulk at fixed angle, strings dominate over loops, so loops were not important for our purposes. It is known \cite{ACV} that at very small angles the contribution from eikonal diagrams is dominant over stringy corrections, so it would be interesting to understand the bulk point singularity in the small angle limit. The methods of \cite{eikonal1,eikonal2} might be helpful for this purpose.\\
\indent The discussion in this work was restricted to zero temperature, but it would be interesting to develop a Mellin representation of correlation functions at finite temperature, and to ask whether the bounds considered here generalize to finite temperature. Relatedly, the nature of the bulk point singularity could be analyzed in AdS/Schwarzschild. This is currently work in progress. 
\appendix

\acknowledgments

We thank C.\ Cardona, J.\ Maldacena, J.\ Penedones, J. Silva, D.\ Simmons-Duffin, and A. Zhiboedov for discussion.  
This work is supported in part by the World Premier International Research Center Initiative,
MEXT, Japan.
The work of H.O. is also supported in part by
U.S.\ Department of Energy grant DE-SC0011632,
by JSPS Grant-in-Aid for Scientific Research C-26400240,
and by JSPS Grant-in-Aid for Scientific Research on Innovative Areas
15H05895.
HO thanks the Aspen Center for Theoretical Physics, which is supported by
the National Science Foundation grant PHY-1607611,  where part of this work was done.


\end{document}